\documentclass{aa}

\usepackage{txfonts}
\usepackage{graphicx}

\newcommand{\pd}[2]{\frac{\partial #1}{\partial #2}}

\begin{document}

\title{Two-component jet simulations}
\subtitle{II. Combining analytical disk and stellar MHD outflow solutions}

\author{
  T. Matsakos  \inst{1} \and
  S. Massaglia \inst{1} \and
  E. Trussoni  \inst{2} \and
  K. Tsinganos \inst{3} \and
  N. Vlahakis  \inst{3} \and
  C. Sauty     \inst{4} \and
  A. Mignone   \inst{1}
}

\authorrunning{Matsakos et al.}
\titlerunning{II. Combining analytical disk and stellar MHD outflow solutions}

\institute{
  Dipartimento di Fisica Generale, Universit\`a degli Studi di Torino,
  via Pietro Giuria 1, 10125 Torino, Italy \\
  \email{matsakos@ph.unito.it} \and
  INAF/Osservatorio Astronomico di Torino, via Osservatorio 20,
  10025 Pino Torinese, Italy \and
  IASA and Section of Astrophysics, Astronomy and Mechanics,
  Department of Physics, University of Athens, \\
  Panepistimiopolis, 15784 Zografos, Athens, Greece \and
  Observatoire de Paris, L.U.Th., 92190 Meudon, France
}

\date{Received ?? / Accepted ??}

\abstract{
  Theoretical arguments along with observational data of YSO jets suggest the
  presence of two steady components: a disk wind type outflow needed to explain
  the observed high mass loss rates and a stellar wind type outflow probably
  accounting for the observed stellar spin down.
  Each component's contribution depends on the intrinsic physical properties of
  the YSO-disk system and its evolutionary stage.
}{
  The main goal of this paper is to understand some of the basic features of the
  evolution, interaction and co-existence of the two jet components over a
  parameter space and when time variability is enforced.
}{
  Having studied separately the numerical evolution of each type of the
  complementary disk and stellar analytical wind solutions in Paper I of this
  series, we proceed here to mix together the two models inside the
  computational box.
  The evolution in time is performed with the PLUTO code, investigating the
  dynamics of the two-component jets, the modifications each solution undergoes
  and the potential steady state reached.
}{
  The co-evolution of the two components, indeed, results in final steady state
  configurations with the disk wind effectively collimating the inner stellar
  component.
  The final outcome stays close to the initial solutions, supporting the
  validity of the analytical studies.
  Moreover, a weak shock forms, disconnecting the launching region of both
  outflows with the propagation domain of the two-component jet.
  On the other hand, several cases are being investigated to identify the role
  of each two-component jet parameter.
  Time variability is not found to considerably affect the dynamics, thus making
  all the conclusions robust.
  However, the flow fluctuations generate shocks, whose large scale structures
  have a strong resemblance to observed YSO jet knots.
}{
  Analytical disk and stellar solutions, even sub modified fast ones, provide
  a solid foundation to construct two-component jet models.
  Tuning their physical properties along with the two-component jet parameters
  allows a broad class of realistic scenarios to be addressed.
  The applied flow variability provides very promising perspectives for the
  comparison of the models with observations.
}

\keywords{
  ISM/Stars: jets and outflows -- MHD -- Stars: pre-main sequence, formation
}

\maketitle

\section{Introduction}
  \label{sec:introduction}

Jets are supersonic and highly collimated plasma outflows emanating from a
plethora of astrophysical objects.
In particular, those associated with Young Stellar Objects (YSO) have been found
to be accretion powered (Cabrit et al.\cite{Cab90}; Hartigan et al.
\cite{Har95}), to have narrow opening angles and to propagate for several
hundreds of AU (Dougados et al. \cite{Dou00}; Hartigan et al. \cite{Har04}).
Although their large scale properties are rather well known, the conditions at
the launch regions are still unclear.
The new generation of high angular resolution instrumentation is expected to
adequately resolve the central regions of YSOs and hence constrain the various
theoretical models that currently exist.

A promising scenario supported by both observational data and theoretical
arguments is that of a two-component jet, wherein a pressure driven stellar
outflow is surrounded by a disk wind.
In particular, \ion{He}{i} $\lambda 10830$ profiles of classical T Tauri
stars (CTTS) indicate the presence of two genres of wind (Edwards et al.
\cite{Edw06}; Kwan et al. \cite{Kwa07}).
One is ejected radially with respect to the central object and the other is
launched at a constant angle with respect to the equatorial plane.
As a result, CTTS may be classified according to their outflow properties.
Some of them seem to be associated with a stellar origin, others with a disk
origin and the rest with both components having roughly equivalent
contributions.
Therefore, it is suggested that both types of winds participate, with the
dominance being dictated by the intrinsic physical factors of the specific YSO.

Such a scenario (e.g. Sauty \& Tsinganos \cite{Sau94}; Shu et al. \cite{Shu94})
is supported by theoretical arguments as well.
Ferreira et al. (\cite{Fer06}) conclude that YSO jets consist of two types of
steady winds plus a sporadic outflow.
An extended disk wind, which is required for the explanation of the high mass
fluxes observed in optical jets and an inner pressure driven outflow of stellar
origin (Bogovalov \& Tsinganos \cite{Bog01}) collimated by the disk wind.
A third component is expected to be launched due to the variable conditions of
the thin layer between the protostellar magnetosphere and the disk's magnetic
field.
Their interaction may drive weak sporadic mass ejections probably associated
with jet variability.

In favor of the two-component jet scenario, there is also the yet unresolved
question of the protostellar spin down.
Matt \& Pudritz (\cite{Mat05}; \cite{Maa08}; \cite{Mab08}) have shown that the
disk-locking mechanism, which was believed to slow down the rotation of the
central object, is not in good agreement with observations.
On the contrary, they propose that the stellar wind is capable of and most
likely responsible for the spin down of the protostar.
A wide parameter space has been investigated to support such a conclusion,
whereas it is argued that the physical mechanisms which drive the actual
launching are less important, hence allowing all sorts of stellar wind models.

A plethora of studies exists in the literature concerning numerical simulations
performed to investigate the launching and propagation of jets.
Two approaches are adopted: in the one, the disk is treated as a boundary (e.g.
Pudritz et al. \cite{Pud06}; Fendt \cite{Fen06}; and references therein), while
in the other the disk is included in the computational box, hence studying its
dynamics simultaneously and self consistently with those of the jet (first
studied in Casse \& Keppens \cite{Cas02}, \cite{Cas04}).
More recently, Meliani et al. (\cite{Mel06}) effectively incorporated a stellar
type outflow accelerated by turbulent heating and in Meliani \& Keppens
(\cite{Mel07}), the transverse stability of relativistic two-component jets was
examined.
Furthermore, adopting a different initial setup, Zanni et al. (\cite{Zan07})
studied the effects of resistivity on the dynamics of the disk-jet system
and Tzeferacos et al. (\cite{Tze??}) performed an interesting parameter study on
disk magnetization.

Despite the complexity of the non-linear MHD equations, the derivation of
analytical steady state outflow solutions has proved successful in the context
of self-similarity (Vlahakis \& Tsinganos \cite{Vla98}).
Each family of these solutions (radially or meridionally self-similar) manages
to capture the physical mechanisms involved in either disk winds (Blandford \&
Payne \cite{Bla82}; Ferreira \cite{Fer97}; Vlahakis et al. \cite{Vla00},
hereafter VTST00), or stellar outflows (Sauty \& Tsinganos \cite{Sau94};
Trussoni et al. \cite{Tru97}; Sauty et al. \cite{Sau02}, hereafter STT02).
The geometrical properties of these two classes of solutions are complementary.
Although radially self-similar models become singular at small polar angles, the
meridionally self-similar ones are by definition appropriate for modeling of the
outflow at the axis.

In the first paper of this series (Matsakos et al. \cite{Mat08}, Paper I), we
addressed the topological stability, as well as several physical and numerical
properties, separately for typical radially and meridionally self-similar
solutions.
Such analytically derived wind models were defined as ADO (Analytical Disk
Outflow) and ASO (Analytical Stellar Outflow), respectively.

Concerning the ADO model, its main feature is the formation of a shock in the
super fast magnetosonic region.
Upstream of this shock, the analytical and the asymptotic numerical solutions
are basically coincident, while the downstream flow converges to a consistent
physical solution, overcoming the singularity of the analytical model at the
symmetry axis (first achieved in Gracia et al. \cite{Gra06}).
This shock corresponds to the numerically modified fast magnetosonic separatrix
surface (FMSS, Tsinganos et al. \cite{Tsi96}) that causally disconnects the
downstream flow from its launching region.
This property is quite robust to variation of the physical parameters, and has
been recently confirmed also in Stute et al. (\cite{Stu08}), where an outer
radial truncation of the disk wind was imposed in the simulations.
Moreover, a particular model was initialized by specifying a sub modified fast
solution both at the initial conditions and at the boundaries (i.e. a flow
causally connected throughout the whole computational domain).
Over time, the shock was still found, with its position marking the FMSS that
causally separates the upstream and downstream regions.

On the contrary, the ASO model does not show singularities at its boundaries
and therefore, the evolution of its super Alfv\'enic region does not show any
readjustments.
However, since energy input is a vital constituent of the model's acceleration,
the modifications of the energy source terms in the sub Alfv\'enic domain were
demonstrated to strongly affect the outcome of the flow.
In particular, we verified that an adiabatic evolution resulted in a collapse of
the jet to an almost static atmosphere, whereas specifying a polytropic index to
mimic almost isothermal conditions produces a weak collimated turbulent wind.

The goal of the present work is to study the two-component jet scenario, taking
advantage of both analytical and numerical approaches.  
Specifically, we construct models by properly defining the initial conditions
with a mixture of two analytical (ADO \& ASO) solutions connected through a
transition region.
The introduction of a few normalization and mixing parameters, along with
enforced time variability applied to the stellar component or at the matching
surface, allows the examination of several interesting cases.

The paper is structured as follows.
Section \S\ref{sec:analytical_models} revises a few basic properties of the
analytical solutions, section \S\ref{sec:numerical_models} describes the mixing
procedure followed to set the initial conditions.
In the same section, the different cases investigated are presented along with
the numerical setup.
In section \S\ref{sec:results} we discuss the results of the simulations
performed.
Section \S\ref{sec:conclusions} summarizes and reports the conclusions of this
work.

\section{MHD equations and the analytical solutions}
  \label{sec:analytical_models}

Our starting point is the ideal MHD equations for the conservation of mass,
momentum, energy and magnetic flux together with the flux-freezing condition:
\begin{equation}
  \pd{\rho}{t} + \nabla \cdot (\rho \vec V) = 0\,,
  \label{eq:density}
\end{equation}
\begin{equation}
  \pd{\vec V}{t} + (\vec V \cdot \nabla)\vec V
    + \frac{1}{\rho}\vec B \times (\nabla \times \vec B)
    + \frac{1}{\rho}\nabla P = -\nabla \Phi\,,
\end{equation}
\begin{equation}
  \pd{P}{t} + \vec V \cdot \nabla P
    + \Gamma P \nabla \cdot \vec V = \Lambda\,,
  \label{eq:energy}
\end{equation}
\begin{equation}
  \pd{\vec B}{t} - \nabla \times (\vec V \times \vec B) = 0
    \quad \mathrm{and} \quad \nabla\cdot\vec B = 0\,,
  \label{eq:induction}
\end{equation}
where $\rho$, $P$, $\vec{V}$, $\vec{B}$ are the density, pressure, velocity and
magnetic field (over $\sqrt{4\pi}$), respectively.
The gravitational potential, $\Phi,$ is equal to $-\mathcal{GM}/R$ with
$\mathcal{G}$, $\mathcal{M}$ and $R$ denoting the gravitational constant, the
mass of the central object and the spherical radius, respectively.
$\Lambda$ represents the volumetric energy gain/loss terms
($\Lambda = [\Gamma - 1] \rho \mathcal{Q}$, with $\mathcal{Q}$ the energy source
terms per unit mass), and $\Gamma$ is the ratio of the specific heats.

Assuming steady state conditions and axisymmetry, several conserved quantities
exist along the fieldlines (e.g. Tsinganos \cite{Tsi82}).
These are the mass to magnetic flux ratio $\Psi_A$, the angular velocity of the
footpoints of the fieldlines $\Omega$ and the total angular momentum flux to
mass flux ratio\footnote{
  In Paper I, VTST00, STT02 and all previous studies on self-similar outflows,
  this integral was defined as ``specific angular momentum''.
} $L$. 
If $\Lambda = (\Gamma-\gamma)P\nabla\cdot\vec{V}$, where $\gamma$ is the
polytropic index (see \S\ref{sec:energetics}), also the total energy flux to
mass flux $E$ and the specific entropy $Q$ are conserved along the streamlines.

In the paper we adopt the following notation: subscripts $D$ and $S$ are used to
refer to the ADO and ASO solutions, respectively, while $(r, \phi, z)$ and
$(R, \theta, \phi)$ are the cylindrical and spherical coordinates.
Note that in Paper I the subscript $r$ was used for the ADO model and $\theta$
for the ASO solution.
The subscript $*$ denotes a constant of the order of unity which is used for the
relative normalization of the two solutions in order to correspond, for
instance, to a solution of the same protostellar mass, as will be explained in
\S\ref{sec:numerical_models}.
The values of the starred quantities correspond to the non-dimensional physical
variables at the Alfv\'enic surfaces of each model at the reference fieldline
$\alpha = 1$ (see below).
Finally, subscript zero in a quantity $U_0$ is used to introduce dimensions in
the code units $U$, i.e. $U' = U_0U$, where $U'$ is the physical value of a
variable given in cgs.

\subsection{The analytical models}

We employ the ADO solution which is described in VTST00 and implemented in Paper
I, that successfully crosses all three critical surfaces.
The ASO model we adopt is a solution similar to the one presented in the first
article of this series, taken from STT02, but with different parameter values:
higher mass loss rate, larger magnetic lever arm and a non spherically symmetric
gas pressure.
Here we only provide a few aspects of the analytical solutions, whereas the
model parameters are reported in Table~\ref{tab:adso_parameters}\footnote{
  The value of the parameter $x = 0.75$, is related to the ejection index $\xi$
  of Ferreira (\cite{Fer97}) and corresponds to zero ejection according to its
  expression.
  However, from Figs.~5 and 6 of VTST00 it is evident that the solution with
  $x = 0.7575$, i.e. $\xi = 0.0025$, is almost identical to the one with
  $x = 0.75$ for $z \gtrsim 0.1$.
  Therefore, we argue that the ADO solution employed here should not contradict
  the theoretical arguments presented in Ferreira (\cite{Fer97}).}$^,$\footnote{
  $\lambda$ and $\lambda'$ are related to the rotational velocity, $\kappa$ and
  $\delta$ to the longitudinal profile of the pressure and density,
  respectively, $\epsilon$ to the energetic balance across the poloidal
  fieldlines and $\beta$ to the energy input.
  The constants $\mathcal K$ and $\nu$ measure the gravitational potential for
  each solution, whereas $\mu$ is associated to the relative magnitudes of
  magnetic and thermal pressure.}
and the explicit formulae of the physical variables are provided in Appendix
\ref{app:dependencies}.
Further technical information on the solutions can be found in Paper I and
references therein.

\begin{table}
  \caption{Parameters characterizing the adopted analytical solutions.}
  \label{tab:adso_parameters}
  \centering
    \begin{tabular}{c c c c c c}
    \hline\hline
    \multicolumn{6}{c}{ADO solution}                      \\
    \hline
    $x$  & $\gamma$ & $\lambda$ & $\mu$ & $\mathcal{K}$ & \\
    0.75 & 1.05     & 11.7      & 2.99  & 2.00          & \\
    \hline
         &          &           &       &               & \\
    \hline\hline
    \multicolumn{6}{c}{ASO solution} \\
    \hline
    $\kappa$            & $\beta$ & $\delta$            & $\lambda'$ &
      $\epsilon$         & $\nu$ \\
    $2.10\times10^{-2}$ & 1.00    & $7.78\times10^{-2}$ & 7.75       &
      $1.2\times10^{-2}$ & 1.50 \\
    \hline
  \end{tabular}
\end{table}

Recalling a few useful expressions, the starred quantities for each analytical
model are related in the following manner:
\begin{equation}
  V_{D*} = \frac{B_{D*}}{\sqrt{\rho_{D*}}}\,, \quad
    P_{D*} = \frac{\mu B_{D*}^2}{2}\,, \quad
    \mathcal K = \sqrt{\frac{g}{r_* V_{D*}^2}}\,,
  \label{eq:ado_relations}
\end{equation}
\begin{equation}
  V_{S*} = \frac{B_{S*}}{\sqrt{\rho_{S*}}}\,, \quad
    P_{S*} = \frac{1}{2}B_{S*}^2\,, \quad
    \nu = \sqrt{\frac{2g}{R_* V_{S*}^2}}\,,
  \label{eq:aso_relations}
\end{equation}
where $r_*$ and $R_*$ correspond to the non-dimensional distances of the
Alfv\'enic surfaces of the ADO and ASO solutions, respectively, and $g = 4$ is
the constant of the gravitational force in code units.

The magnetic field of each solution is given by the following formula:
\begin{equation}
  \vec{B} = \frac{1}{r}\nabla A \times \hat{\phi} + B_\phi\hat{\phi}\,.
  \label{eq:bp}
\end{equation}
labels the iso-surfaces that enclose constant poloidal magnetic flux, i.e. the
magnetic fieldlines.
In particular, for the ADO solution, $A$ is given by:
\begin{equation}
  A_D = \frac{B_{D*}r_*^2}{x}\alpha_D^{x/2}\,,
    \quad \mathrm{where} \quad
    \alpha_D = \frac{r^2}{r_*^2G_D^2}\,.
\end{equation}
Similarly, for the ASO model:
\begin{equation}
  A_S = \frac{B_{S*}R_*^2}{2}\alpha_S\,,
    \quad \mathrm{where} \quad
    \alpha_S = \frac{r^2}{R_*^2G_S^2}\,.
\end{equation}
The values of $G_D(\theta)$ and $G_S(R)$ are provided by the analytical
solutions (see Paper I and references therein for more details).

We provide here the measure of the magnetic lever arm, braking the disk or star
for each solution, as defined in Ferreira et al. (\cite{Fer06}).
This is the same for all fieldlines, and is given by the relation
$\lambda \simeq r_{A}^2/r_{fp}^2$:
\begin{equation}
  \lambda_D \simeq \frac{1}{G_D^2(\pi/2)} \simeq 40\,,
\end{equation}
\begin{equation}
  \lambda_S \simeq \frac{1}{G_S^2(R_{bs})} \simeq 330\,.
\end{equation}
where $r_{fp}$ is the cylindrical distance of the footpoint of a particular
fieldline and $r_{A}$ is the cylindrical distance of its Alfv\'enic point.
$G_D(\pi/2)$ and $G_S(R_{bs})$ correspond to the values of the analytical
solutions at the equatorial plane and at the base of the stellar wind, $R_{bs}$,
respectively.

\section{The numerical models}
  \label{sec:numerical_models}

In order to choose physical scales, we set the length, and density code units
equivalent to
$r_0 = 1\,$AU and $\rho_0 = 10^{-12}\,$g$\,$cm$^{-3}$.
In addition we assume the protostar to be of one solar mass,
$\mathcal{M} = 1\,M_{\sun}$.
Then, since the MHD Eqs.~(\ref{eq:density}) - (\ref{eq:induction}) are written
in non dimensional form, it can be easily derived that:
$V_0 = \sqrt{\mathcal{GM}/gr_0} = 14.9\,$km$\,$s$^{-1}$,
$P_0 = \rho_0V_0^2 = 2.22\,$dyne$\,$cm$^{-2}$ and
$B_0 = \sqrt{4\pi P_0} = 5.28\,$G.
Hence, the time unit corresponds to $t_0 = 0.32\,$y.

\subsection{Normalization}

Now, we normalize the solutions to each other by defining the three ratios,
which are parameters of the two-component jet models:
\begin{equation}
  \ell_l = \frac{R_*}{r_*}\,,\quad
  \ell_V = \frac{V_{S*}}{V_{D*}}\,,\quad
  \ell_B = \frac{B_{S*}}{B_{D*}}\,,
\end{equation}
where the subscripts $l$, $V$ and $B$ stand for length, velocity and magnetic
field, respectively.
As it will be seen, only $\ell_B$ can be chosen freely, while the other two are
fixed by physical arguments and the properties of the analytical solutions.
More precisely, observations indicate that the launching region of disk winds
lies in the range $0.2 - 3\,$AU (Bacciotti et al. \cite{Bac02}; Anderson et al.
\cite{And03}; Coffey et al. \cite{Cof04}).
Therefore, demanding that the reference fieldline $\alpha_D = 1$ is rooted at
$0.16\,$AU on the equatorial plane, we find $r_* = 1$.
Moreover, assuming that the region where the stellar wind is being launched is
roughly at $0.01\,$AU or at $R_{bs} = 0.01$ in code units, we derive $R_* = 0.1$
and hence $\ell_l = 0.1$.
It follows from relations (\ref{eq:ado_relations}) and (\ref{eq:aso_relations})
that $V_{D*} = 1$, $V_{S*} = 5.96$, and thus $\ell_V = 5.96$.
Finally, we arbitrarily set $B_{D*} = 1$ and the choice of $B_{S*}$ will control
$\ell_B$.

\subsection{The mixing function}

Since the mixing will depend on the magnetic fieldlines, we define a trial
magnetic flux function by the simple sum: $A_{tr} = A_D + A_S$.
We point out that this quantity will help only in the mixing procedure and will
not be used to generate the magnetic field present in the initial conditions.
We further define the mixing function:
\begin{equation}
  U_{2comp} = w_DU_D + w_SU_S\,,
  \label{eq:mixing_function}
\end{equation}
with the weights $w_D$ and $w_S$ given by:
\begin{equation}
  w_D = 1 - w_S
    \quad \textrm{and} \quad
    w_S = \exp\left[-\left(\frac{A_{tr}}{qA_{m}}\right)^d\right]\,.
\end{equation}
In the latter expressions, $A_{m} = 1.33$ is a constant corresponding to the
matching surface rooted at $0.16r_*$, i.e. at $0.16$AU on the equatorial plane,
$q$ is a parameter that effectively moves this surface closer to the protostar
and $d$ sets the steepness of the transition from the inner ASO to the outer ADO
solution.

The initial values of the physical variables $\rho$, $P$, $V_z$, $V_{\phi}$,
$B_{\phi}$ are set up using relation (\ref{eq:mixing_function}).
Moreover, with the help of the same expression, we initialize the two-component
magnetic flux function\footnote{
  Note that $A_{tr}$ should not be confused with $A$.
  Although the former is a simple sum of $A_D$ and $A_S$, the latter is computed
  from the mixing function, as the rest of the variables.
} $A$, from which the poloidal component of the magnetic field is generated
using Eq.~(\ref{eq:bp}).
Finally, $V_r$ is initialized following the ideal MHD condition, i.e. demanding
that the poloidal magnetic field is parallel to the poloidal velocity:
\begin{equation}
  V_r = \frac{V_zB_r}{B_z}\,.
\end{equation}

Essentially, such a mixing function provides an exponential damping of each
solution around a particular fieldline of the combined magnetic field.
Therefore, close to the axis, the ASO model dominates, whereas the ADO becomes
the main contributor at the outer regions.

The two-component jet numerical models can be constructed by specifying the
three normalization parameters, $\ell_l$, $\ell_V$, $\ell_B$, and the three
mixing parameters, $A_m$, $q$, $d$.
As it has already been explained, $\ell_l$, $\ell_V$ and $A_m$ are given a fixed
value, leaving $\ell_B$, $q$ and $d$ free to examine a variety of two-component
scenarios.
The latter three parameters control the respective dominance, the location of
the matching surface in between the protostar-disk region and the steepness of
the transition region.
Different values in this parameter space may address the various T Tauri outflow
types and their evolutionary stage.
One would expect that in many cases the efficiency of disk winds would manifest
the early phases of the YSO-disk system, whereas stellar winds would eventually
dominate, especially after the disk has accreted and during the arrival of the
star on the main sequence.

\subsection{Time variability}

Accretion, which controls the conditions at the base of stellar winds, is not
steady in time but rather varies over different time scales ranging from hours,
to days, months, even years (Alencar \& Batalha \cite{Ale02}; Stempels \&
Piskunov \cite{Ste02}; Johns \& Basri \cite{Joh95}).
On the other hand, the protostar is expected to show some sort of variability as
well, for instance the phenomenon of the $11\,$yr solar cycle.
Therefore, the introduction of time variability in the inner stellar component
will allow us to study the stability issues of more general and realistic
scenarios.
In order to achieve this we prescribe the following function:
\begin{equation}
  f_S(r, t) = 1 + \frac{1}{2}\sin\left(\frac{2\pi t}{T_{var}}\right)
    \exp\left[-\left(\frac{r}{2r_m}\right)^2\right]\,,
  \label{eq:stellar_perturbation}
\end{equation}
where $T_{var}$ is the period of the pulsation and $r_m = 5$ is roughly the
cylindrical radius at which the matching separatrix intersects the lower
boundary $z = 10$ of the computational box.
We enforce a sinusoidal time variability depending on $T_{var}$, by multiplying
a physical quantity of the lower boundary with $f(t)$.
The exponential in Eq.~(\ref{eq:stellar_perturbation}) helps to contain the
perturbation only at the inner regions, i.e. the stellar component.
Note that flow fluctuations induce the formation of knot-like structures.
Therefore, also including radiation cooling during the evolution (Tesileanu et
al. \cite{Tes08}) would allow a direct comparison with observational data.
However, this is left to a future article of this series.

Since it is believed that a sporadic outflow is driven by the star-disk magnetic
interaction (Ferreira et al. \cite{Fer00}; Matt et al. \cite{Mat02}), we examine
such cases as well.
In this case we adopt a similar function:
\begin{equation}
  f_X(r, t) = 1 + \frac{1}{2}\sin\left(\frac{2\pi t}{T_{var}}\right)
    \exp\left[-\left(\frac{r - r_m}{r_m}\right)^2\right]\,.
  \label{eq:x_point_perturbation}
\end{equation}

\subsection{Energetics}
  \label{sec:energetics}

We set $\Lambda = (\Gamma - \gamma)P(\nabla\cdot\vec V)$ in
Eq.~(\ref{eq:energy}) with $\gamma = 1.05$.
This assumption, originally made for the derivation of the ADO solution, is
equivalent to a polytropic relation $P \propto \rho^\gamma$ along each
fieldline.
Essentially, it represents the adiabatic evolution of a gas with a ratio of
specific heats $\gamma$, which corresponds to the following energy conservation
law that is solved over time:
\begin{equation}
  \pd{P}{t} + \vec V \cdot \nabla P + \gamma P \nabla \cdot \vec V = 0 \,.
\end{equation}

Recall that in Paper I, simulations were carried out both for the ADO and the
super Alfvenic regions of the ASO solution to test the effects of such an
energetic assumption ($\gamma = 1.05$), as well as an isothermal
($\gamma = 1.0$) or an adiabatic one ($\gamma = \Gamma = 5/3$).
For each model, these different cases produced almost identical results, thus
allowing us to safely adopt this simplification of the energy equation.

\subsection{The numerical two-component jet models}

\begin{table*}
  \caption{
    A short description and the parameters of the unperturbed numerical models.
    The non listed parameters are common for all models and have the following
    values: $\ell_l = 0.1$, $\ell_V = 5.96$ and $A_m = 1.33$.
  }
  \label{tab:unperturbed_models}
  \centering
  \begin{tabular}{l c c c l}
    \hline
    \hline
    Name   & $\ell_B$ & $q$ & $d$ &
    Description \\
    \hline
    1-q01  & 1.0      & 0.1 & 2.0 &
    Small ASO contribution, matching very close to protostar
      (Fig.~\ref{fig:matching_surface})                         \\
    2-q02  & 1.0      & 0.2 & 2.0 &
    Small ASO contribution, matching close to protostar         \\
    3-q05  & 1.0      & 0.5 & 2.0 &
    Small ASO contribution, matching close to disk
      (Fig.~\ref{fig:matching_surface})                         \\
    4-q01  & 2.0      & 0.1 & 2.0 &
    Medium ASO contribution, matching very close to protostar
      (Fig.~\ref{fig:matching_surface})                         \\
    5-q02  & 2.0      & 0.2 & 2.0 &
    Medium ASO contribution, matching close to protostar
      (Figs.~\ref{fig:time_evolution}, \ref{fig:time_difference},
      \ref{fig:critical_surfaces}, \ref{fig:variables},
      \ref{fig:integrals}, \ref{fig:characteristics})           \\
    6-q05  & 2.0      & 0.5 & 2.0 &
    Medium ASO contribution, matching close to disk
      (Fig.~\ref{fig:matching_surface})                         \\
    7-B05  & 0.5      & 0.2 & 2.0 &
    Very small ASO contribution
      (Fig.~\ref{fig:lambda_b})                                 \\
    8-B5   & 5.0      & 0.2 & 2.0 &
    Large ASO contribution
      (Fig.~\ref{fig:lambda_b})                                 \\
    9-B10  & 10.0     & 0.2 & 2.0 &
    Very large ASO contribution                                 \\
    10-d1  & 2.0      & 0.2 & 1.0 &
    Medium ASO contribution, smooth transition
      (Fig.~\ref{fig:steepness})                                \\
    11-d4  & 2.0      & 0.2 & 4.0 &
    Medium ASO contribution, steep transition
      (Fig.~\ref{fig:steepness})                                \\
    \hline
  \end{tabular}
\end{table*}

\begin{table*}
  \caption{
    The time variable numerical models. The two-component jet parameters are the
    same for all cases: $\ell_l = 0.1$, $\ell_V = 5.96$, $\ell_B = 2.0$,
    $A_m = 1.33$, $q = 0.2$ and $d = 2.0$.
  }
  \label{tab:time_variable_models}
  \centering
  \begin{tabular}{l c c c l}
    \hline
    \hline
    Name & $T_{var}/T_K$ & Variability & Variable wind & Description \\
    \hline
    1-S1      & 1      & $V_z$           & Stellar &
      Very high frequency velocity fluctuations of the stellar component
      (Fig.~\ref{fig:aso_variability}) \\
    2-S10     & 10     & $V_z$           & Stellar &
      High frequency velocity fluctuations of the stellar component \\
    3-S10$^2$ & $10^2$ & $V_z$           & Stellar &
      Medium frequency velocity fluctuations of the stellar component
      (Figs.~\ref{fig:aso_variability}, \ref{fig:crit_surf_var}) \\
    4a-S10$^3$ & $10^3$ & $V_z$           & Stellar &
      Low frequency velocity fluctuation of the stellar component 
      (Fig.~\ref{fig:low_frequency}) \\
    4b-S10$^3$ & $10^3$ & $V_z$           & Stellar &
      Low freq. vel. fluct. (lower magnitude: $\pm20\%$) of the stellar
      component (Fig.~\ref{fig:low_frequency}) \\
    5-S10$^4$ & $10^4$ & $V_z$           & Stellar &
      Very low frequency velocity fluctuations of the stellar component
      (Fig.~\ref{fig:low_frequency}) \\
    6-X1      & 1      & $V_z\,\&\,\rho$ & X-type  &
      Very high frequency momentum fluctuations around the X-point
      (Fig.~\ref{fig:x_variability}) \\
    7-X10     & 10     & $V_z\,\&\,\rho$ & X-type  &
      High frequency momentum fluctuations around the X-point \\
    8-X10$^2$ & 10$^2$ & $V_z\,\&\,\rho$ & X-type  &
      Medium frequency momentum fluctuations around the X-point
      (Fig.~\ref{fig:x_variability}) \\
    \hline
  \end{tabular}
\end{table*}

Table~\ref{tab:unperturbed_models} lists the unperturbed numerical two-component
jet models along with their parameters.
Table~\ref{tab:time_variable_models} presents those constructed to investigate
the stability and structure when time variability is applied at the stellar wind
or at the matching surface, effectively mimicking an X-type wind.
In particular, the second column of Table~\ref{tab:time_variable_models} reports
the ratio of the periodicity of the enforced fluctuation, $T_{var}$, over the
Keplerian rotation period $T_K$ ($\sim0.4\,$days) of the protostellar radius,
located roughly at $0.01\,$AU.
This means that we address phenomena with time scales associated with accretion
and the physical conditions present at the star-disk region.
Note that in our models, the Keplerian period of the equatorial footpoint of the
matching surface is of the order of $10\,$days.
In the third and fourth columns of Table~\ref{tab:unperturbed_models} we
indicate the physical quantity that is varied and where it is perturbed,
respectively (i.e. adopting Eq.~[\ref{eq:stellar_perturbation}] or
[\ref{eq:x_point_perturbation}]).

\subsection{PLUTO code and the numerical setup}

The simulations are performed with PLUTO\footnote{
  Publicly available at \texttt{http://plutocode.to.astro.it}}
(Mignone et al. \cite{Mig07}), a versatile shock-capturing numerical code
suitable for the solution of high-mach number flows.
The grid is set up in axisymmetric cylindrical coordinates (2.5D), leaving the
study of azimuthal stability for a future work.
Second order accuracy is applied in both space and time, and the Lax-Friedrichs
solver is adopted.
However, the choice of a particular solver was not found to influence the
results.
The $\nabla\cdot\vec B = 0$ condition is ensured with the 8-wave formulation.

The length code unit is equivalent to $1\,$AU, and therefore the correspondence
to real physical distances is straightforward.
We consider a computational box with $0 \le r \le 100$ and $10 \le z \le 210$
for the unperturbed models, and with $0 \le r \le 50$ and $10 \le z \le 110$
for the time variable ones, omitting the acceleration region of the ASO
solution.
There are two reasons for doing so.
The first argument concerns the complexities appearing when the ADO solution is
initialized in a computational domain that approaches the origin in cylindrical
coordinate systems (as demonstrated in Paper I).
Second, the ASO solution provides a time independent energy source term, which
if included, will artificially constrain time evolution, as shown in Paper I.
In addition, the complicated processes of the ejection and acceleration of
stellar winds are as yet unresolved and hence it is better to first address the
simpler dynamics of two-component jets with the stellar outflow already being
super Alfv\'enic.
Besides, the launching of each component takes place at different and extended
locations of the YSO-disk system and therefore the interaction happens at higher
altitudes.
Moreover, the low frequency models, 4a-S10$^3$, 4b-S10$^3$ and 5-S10$^5$, are
obviously associated with larger length scales and therefore the vertical
direction is chosen $10 \le z \le 610$ for the former and $10 \le z \le 1210$
for the latter.
Essentially, we address the length scales of a few tens AU radially to a
thousand AU vertically.

All models have a uniform resolution of 256 zones for every $100\,$AU.
However, we have evolved a typical model, 5-q02, also in a finer grid of
$512\times1024$ to investigate the properties common to all models, such as time
evolution features, potential steady states, deviations from analytical
solutions and shock formation.
Nevertheless, the cell size was not found to affect the outcome of the numerical
evolution, a feature of the self-similar models that is also supported by Paper
I.
Furthermore, the unperturbed simulations have been carried out up to a final
time of $80\,$y, equivalent to $8\times10^4T_K$, i.e. $80,000$ Keplerian
rotations of the protostar, or for up to $t = 250$ in code units.
The time unit in the code corresponds to $0.32\,$y.
Due to the greater length and time scales involved in models 4a-S10$^3$,
4b-S10$^3$ and 5S10$^4$ the simulations were run up to a final time of $160\,$y.

At the lower boundary, we keep all variables fixed to their initial values after
the mixing, in agreement with the conclusions of Paper I (wherein a detailed
discussion of the correct treatment of the boundary conditions can be found).
Outflow or extrapolated boundary conditions in the region where the flow enters
the computational domain might artificially influence the long term simulations.
At the axis we apply axisymmetric boundary conditions, whereas at the upper and
right boundaries, we apply outflow conditions.
Note that setting the derivative of $B_\phi$ equal to zero at the right boundary
could cause artificial collimation.
However, the ADO solution dominates at this boundary both in the initial
conditions and over time (as will be seen in the next section).
Therefore, recalling from Paper I that the ADO model maintains its exact
equilibrium in the rightmost regions despite the specification of outflow
conditions, we argue that the configurations studied here are not prone to such
a numerical enforcement.

\section{Results}
  \label{sec:results}

We outline first the results obtained that are common to all two-component jet
simulations and then we discuss the effects of the mixing parameters and the
time variability.

\subsection{Time evolution and steady state}
  \label{sec:time_evolution}

\begin{figure*}
  \resizebox{\hsize}{!}{\includegraphics{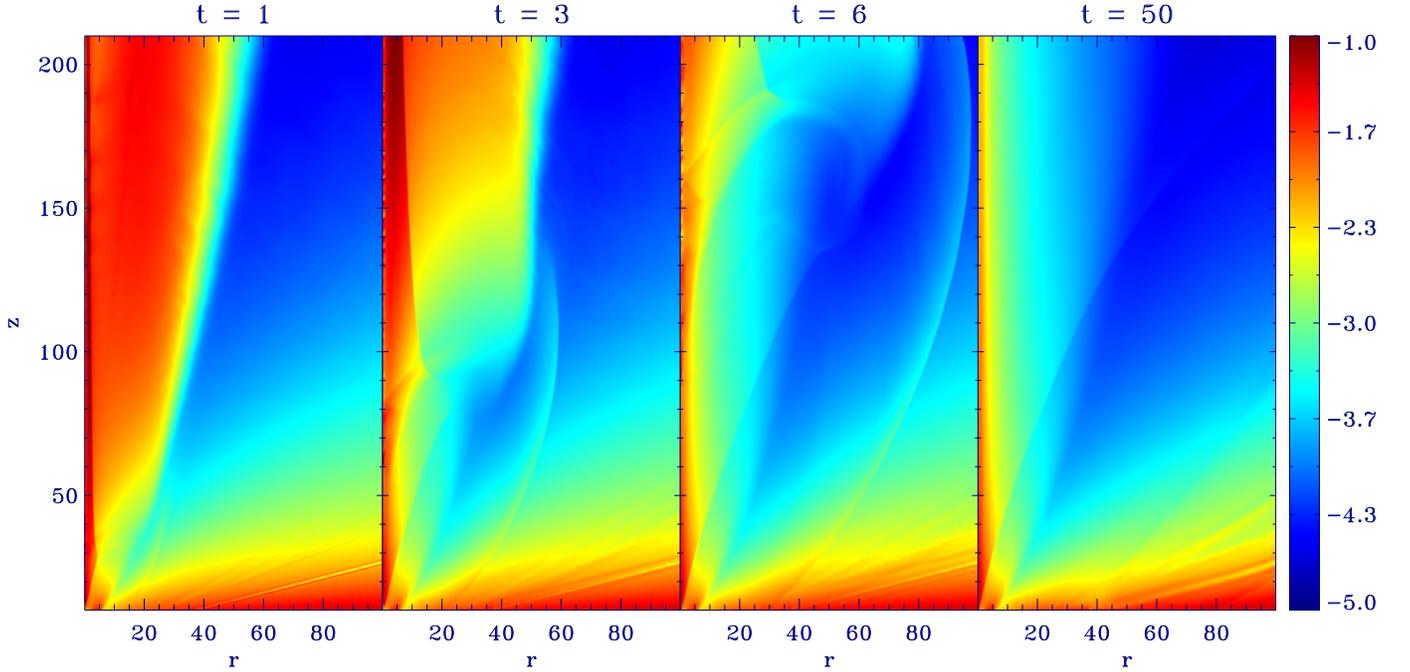}}
  \caption{
    Logarithms of the density at different times for a typical model (5-q02).
    The time unit is $0.32\,$y (or $10$ Keplerian rotations of the footpoint of
    the matching surface).
    In the rightmost plot, a weak shock is observed along the diagonal.
    In addition, the initial matching surface, approximately at $r = 5$ on the
    lower boundary, is still evident.
    To check this, compare it with the leftmost plot which describes a
    configuration very close to the initial one.
  }
  \label{fig:time_evolution}
\end{figure*}

The logarithm of the density is plotted in Fig.~\ref{fig:time_evolution} for
different evolutionary stages of a typical two-component jet model (5-q02).
The initial conditions correspond to equilibrium in the regions where each
analytical solution dominates.
However, around the matching surface, the models are modified and hence a strong
perturbation is generated during the first timesteps of the simulation.
An MHD wave propagates through the ADO solution without leaving behind any
significant rearrangements.
On the contrary, the equilibrium of the ASO model is substantially restructured,
with its density dropping roughly by an order of magnitude.
In only $\sim100$ Keplerian rotations of the footpoint of the matching surface,
the stellar component has already been totally and self consistently modified in
the presence of the ADO solution.

From the rightmost plot of Fig.~\ref{fig:time_evolution}, notice that the
initial matching surface is still evident.
Indeed, this is expected, due to the fixed boundary conditions at the lower
boundary.
At $t = 50$, the initial perturbation has almost left the domain, with the
two-component jet having reached a steady state.
Notice the formation of a weak steady shock, which can be seen almost along the
diagonal direction of the computational domain.

\begin{figure}
  \resizebox{\hsize}{!}{\includegraphics{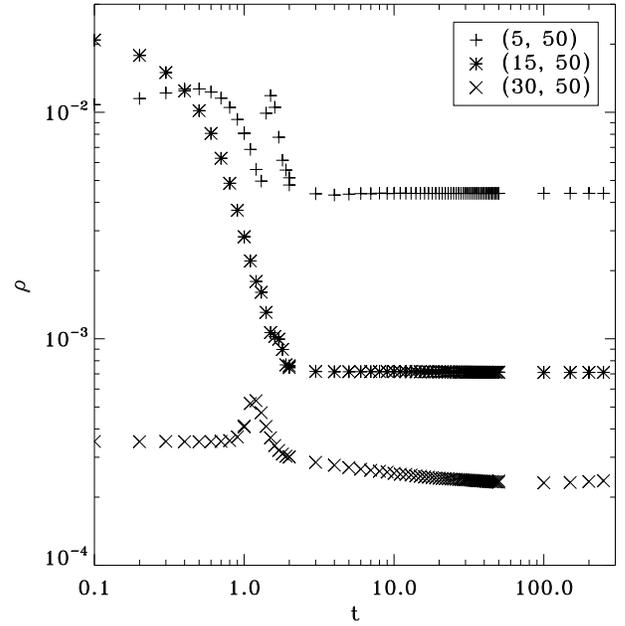}}
  \caption{
    The density fluctuations as a function of time calculated at the points
    $(r_0\,,z_0) = (5\,,50)$, $(15\,,50)$ and $(30\,,50)$.
    The first is located in the ASO dominated region, the second close to the
    matching surface upstream of the shock and the third in the ADO dominated
    region (model 5-q02).
  }
  \label{fig:time_difference}
\end{figure}

In order to establish the conclusion that the two components can co-exist in a
steady state, we plot in Fig.~\ref{fig:time_difference} the density fluctuations
for different time scales of three specific points.
One is located inside the stellar component and the other two upstream of the
shock, at the matching surface and in the disk wind, respectively.
Evidently, for $t > 2.5$ the solution remains almost unchanged up to $t = 250$,
i.e. a time longer by two orders of magnitude.
The disk wind reaches the final exact equilibrium a bit later (at $t\sim25$),
due to the slower wave velocities of this region.
Therefore, the rightmost plot of Fig.~\ref{fig:time_evolution} represents a very
well preserved steady state of the two intrinsically different jet components.

Both the physical and geometrical properties of the two winds are by definition
considerably diverse, since their self-similar symmetries are orthogonal to each
other.
In turn, the same holds true for their respective poloidal critical surfaces.
Therefore, the steady state of such a two-component jet was not a
straightforward expectation.
Nevertheless, it is clearly shown in Figs.~\ref{fig:time_evolution} and
\ref{fig:time_difference} that the two complementary winds manage to co-exist.
Also taking into account the artificial boundary effects present in long term
simulations, investigated in Paper I, the above results are adequate to argue
that the two-component jet models reach a well defined steady state.

Despite the fact that the plots concern a particular model, the same conclusions
are valid for all the unperturbed scenarios presented in
Table~\ref{tab:unperturbed_models}.

\subsection{Deviations from the analytical solutions}
  \label{sec:deviations}

\begin{figure}
  \resizebox{\hsize}{!}{\includegraphics{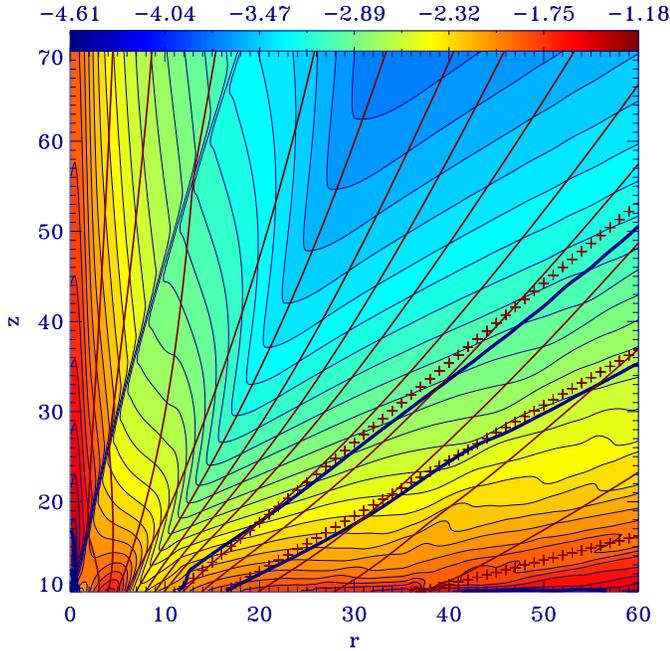}}
  \caption{
    Logarithmic density contours (thin blue lines) for model 5-q02 at $t = 50$.
    The magnetic poloidal fieldlines are overplotted with red lines.
    In the lower right part, going clockwise, the fast magnetosonic, the
    Alfv\'enic and the slow magnetosonic critical surfaces are plotted with
    red crosses for the ADO solution and with thick blue lines for the final
    numerical two-component flow.
  }
  \label{fig:critical_surfaces}
\end{figure}

\begin{figure}
  \resizebox{\hsize}{!}{\includegraphics{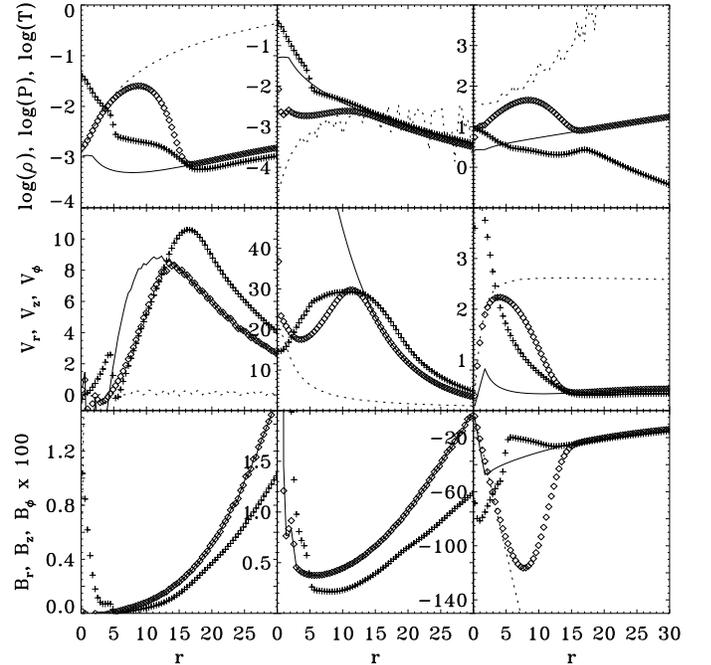}}
  \caption{
    The physical variables plotted at $z = 50$ for the ADO model alone (solid
    line), the ASO model alone (dashed line), the initial setup of 5-q02
    (diamonds) and its final configuration (crosses).
    The quantities displayed from left to right are: in the top row $\log\rho$,
    $\log P$ and $\log T$, in the middle row $V_r$, $V_z$ and $V_\phi$, in the
    bottom row $B_r$, $B_z$ and $B_\phi$.
  }
  \label{fig:variables}
\end{figure}

\begin{figure}
  \resizebox{\hsize}{!}{\includegraphics{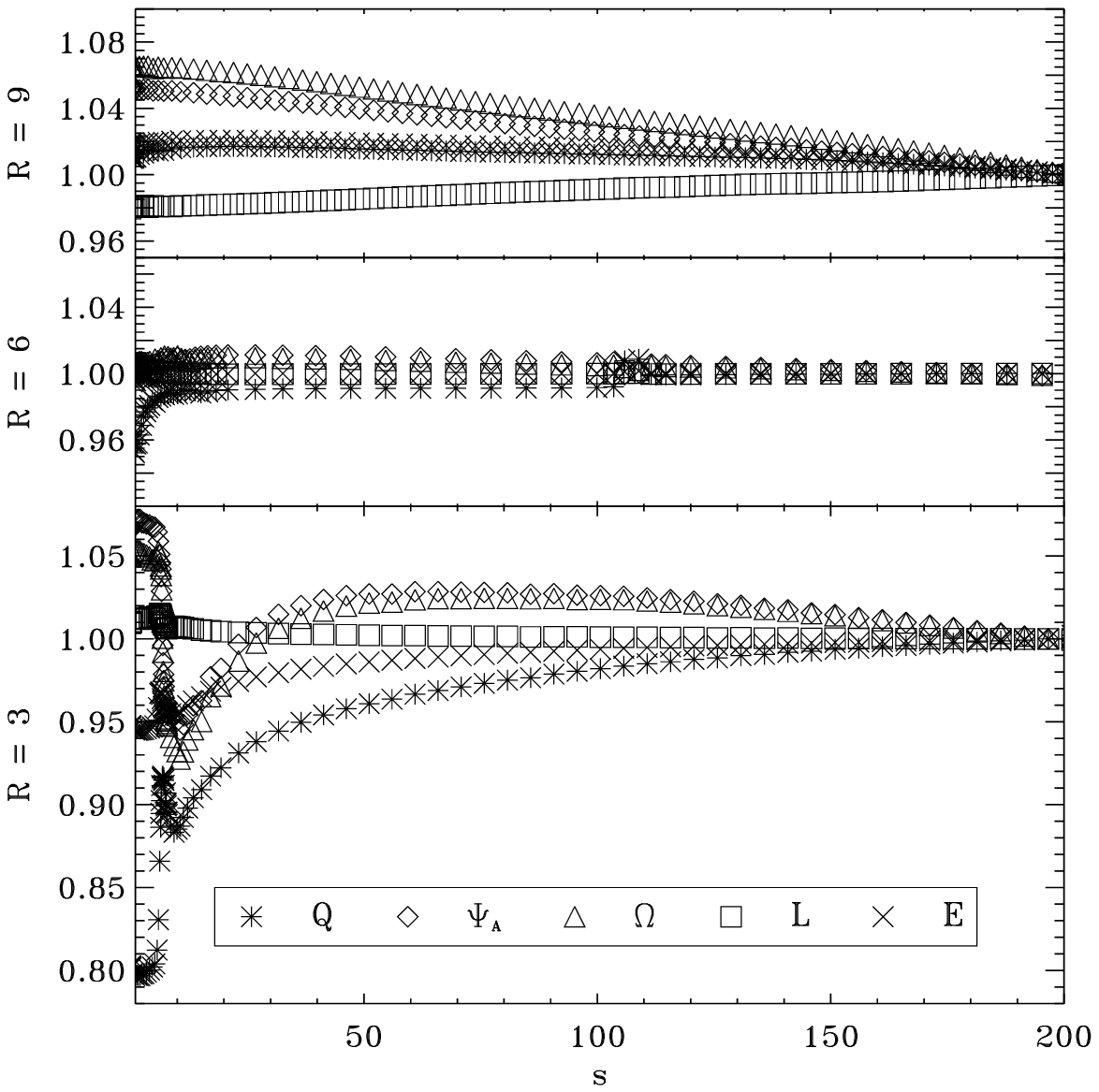}}
  \caption{
    The normalized integrals of motion plotted along three fieldlines of model
    5-q02, rooted at the positions (from bottom to top) $(3\,, 10)$,
    $(6\,, 10)$, and $(9\,, 10)$, corresponding to the ASO, mixing and ADO
    dominated regions, respectively.
    The distance from the lower boundary is parametrized by $s$.
  }
  \label{fig:integrals}
\end{figure}

Another crucial question that arises concerns how close the final outcome of the
simulations is to the initial analytical solutions.
In particular, the smaller the deviations are found to be, the more valid and
robust are the analytical studies on the self-similar MHD outflows.
This also implies the easy and appropriate extension of their conclusions to the
two-component jet scenario, especially for the analytically derived disk winds.

Therefore, in Fig.~\ref{fig:critical_surfaces} we plot the critical surfaces of
the ADO solution (red crosses) and those of the final numerical two-component
jet (thick blue lines), along with the logarithmic density contours and the
magnetic fieldlines (red lines).
The propagation of the perturbation described in \S\ref{sec:time_evolution}
results in the slight modification of the fast magnetosonic and Alfv\'enic
critical surface, as can be seen from their almost perfect match in
Fig.~\ref{fig:critical_surfaces}.
On the other hand, the slow magnetosonic critical surface seems to have
collapsed towards the lower boundary.
The slow waves generated initially at the matching surface by definition cannot
pass to the sub slow domain.
Consequently, if the surface stayed at its initial location, the waves would not
have been able to leave the lower right region of the computational box.
Such a phenomenon is observed in SC3 and SC5 runs of Stute et al.
(\cite{Stu08}), where matter accumulates downstream of the slow critical
surface.
However, in our case, the separatrix is being bent downward, tangentially to the
lower boundary, hence allowing the initial perturbation to exit the simulated
box.
As a result, the lower right region shows a significantly higher degree of
deviation from the initial conditions than the rest of the domain.
Nonetheless, it does reach a steady state asymptotically.
Note that the critical surface cannot be dragged away, due to the fixed boundary
conditions, which describe a sub slow flow at $z = 10$.
Similar features are also observed in Gracia et al. (\cite{Gra06}) and in most
runs of Stute et al. (\cite{Stu08}).

The fact that the system finds an equilibrium so close to the analytical
solution is due to the topological stability of the ADO model discussed in Paper
I.
No matter if the central part of the disk wind is substituted as a whole with a
physically and geometrically different kind of outflow, the solution maintains
all its properties, proving its stability.

The poloidal critical surfaces plotted in Fig.~\ref{fig:critical_surfaces}
also provide other insights into the two-component scenario.
Close to the axis, they have an elliptical shape, as can be seen in the region
very close to (0, 10), and eventually become conical, after the matching
surface.
Intuitively, this makes sense due to the difference in the symmetry of the
accelerating mechanisms of the two winds.
In Kwan et al. (\cite{Kwa07}), two types of outflows are observed, one emanating
radially out of the protostar and the other being ejected at a constant angle
with respect to the disk midplane.
This implies a geometry of the poloidal critical surfaces similar to
Fig.~\ref{fig:critical_surfaces}.

In Fig.~\ref{fig:variables}, the physical variables are plotted at the constant
height $z = 50$, for the initial setup (diamonds) and final configuration
(crosses) of model 5-q02.
In addition, the initial ADO (solid lines) and ASO (dashed lines) solutions are
also shown before their combination.
All plots present the effect of the mixing function.
Close to the axis, the ASO model dominates, whereas approaching the right
boundary, the ADO becomes the main contributor.
A jump can be observed in most quantities, which represents the weak shock
discussed in \S\ref{sec:shock_formation}.
Apart from the density and the poloidal component of the magnetic field, the
initial and final configurations converge at large distances, showing the
stability of the ADO solution.
However, this happens far from the slow magnetosonic critical surface.
The modifications the initial ASO solution undergoes can be seen from the final
equilibrium reached close to the axis.
Note that the temperature plot can be used as a guide when looking for
two-component jet parameters appropriate to address observed jets.

Finally, the normalized integrals of motion (Eqs. [5]-[9] of Paper I) are
plotted in Fig.~\ref{fig:integrals} along three selected fieldlines rooted at
the points $(3\,, 10)$, $(6\,, 10)$, and $(9\,, 10)$.
One is in the ASO dominated part, one is in the ADO domain and the other is
almost along the matching surface crossing the shock.
In all cases the integrals are conserved with high accuracy, varying only within
a few $\%$.
At large distances from the shock, they tend to become constant, which indicates
that the system reaches a steady state in all three regions.
For the two inner fieldlines, at $s = 5$ and $s = 100$, respectively, the
observed jumps are related to the crossing of the shock.
In particular, the larger deviation from constancy occurs for the specific
entropy integral $Q$, as expected.

\subsection{Shock formation}
  \label{sec:shock_formation}

\begin{figure}
  \resizebox{\hsize}{!}{\includegraphics{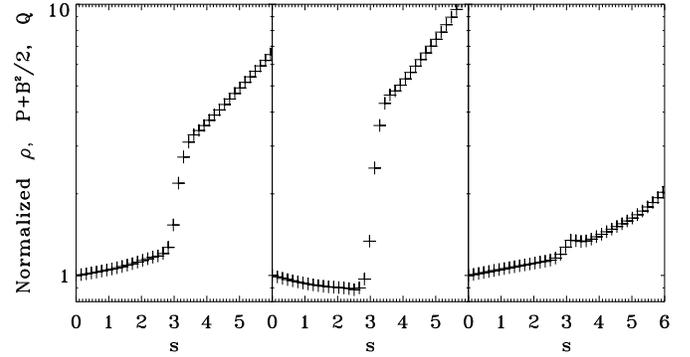}}
  \caption{
    Normalized discontinuities (from left to right) of the density, total
    pressure (thermal plus magnetic) and the entropy $Q$) across the shock,
    close to the point $(5\,, 30)$.
    Notice that $s$ is increasing in the inverse direction of $r$.
  }
  \label{fig:shock}
\end{figure}

\begin{figure}
  \resizebox{\hsize}{!}{\includegraphics{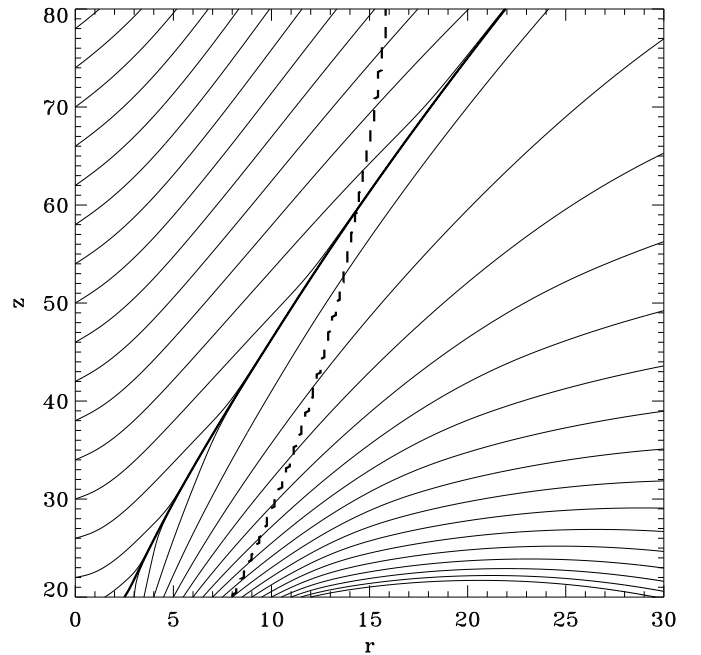}}
  \caption{
    A family of the characteristics (thin solid lines) of the fast magnetosonic
    waves in a zoomed super fast magnetosonic region around the shock (thick
    solid line) for model 5-q02.
    The thick dashed line is the initial matching surface.
  }
  \label{fig:characteristics}
\end{figure}

In Fig.~\ref{fig:shock}, we plot the normalized density, total pressure
($P + B^2/2$) and entropy $Q$ across the shock (direction right to left) around
the position $(5\,, 30)$, very close to the one assumed in Paper I.
This point is located inside the domain where the stellar outflow dominates.
Apparently, the density seems to increase by a factor of 2, whereas the pressure
increases by a factor of 4.
On the contrary, the jump seen in the entropy is very weak, being an order of
magnitude smaller: this is not surprising, recalling that $\gamma = 1.05$, i.e.
conditions very close to isothermal, wherein entropy remains unchanged across
shocks.
Also there is no heating/cooling present in any simulation of this paper, thus
making the above analysis simpler.

In Fig.~\ref{fig:characteristics} we plot one of the two families of the
characteristics of the fast magnetosonic waves (thin lines) for model 5-q02,
along with the initial matching surface (dashed line).
It is evident that the shock (thick solid line) is not crossed by the downstream
characteristics.
This shows the causal disconnection of the two domains, upstream and downstream
of the shock.
In other words, the shock represents the horizon for the propagation of all MHD
waves, coinciding with the numerical FMSS.

This feature is closely related to the ADO solution and was studied in detail
in Paper I.
However, the two-component case we present here is especially interesting for
the following reason.
The shock manifests even in the central area, where the contribution of the ASO
model is total.
This implies that it is not associated with the lower boundary, but on the
contrary, it forms above it, intersecting the symmetry axis.
Taking also into account the results of Paper I, the shock seems to be an
intrinsic feature of the ADO solution.
Consequently, the presence of the disk wind model in the two-component jet
scenario has the remarkable characteristic of producing outflows that are
causally disconnected to their launching region, despite the fact that the
initial conditions causally connect the whole computational box.

\subsection{Parameter study}

In this section, we present the behavior of the two-component jets when we
change the model parameters.

\begin{figure}
  \resizebox{\hsize}{!}{\includegraphics{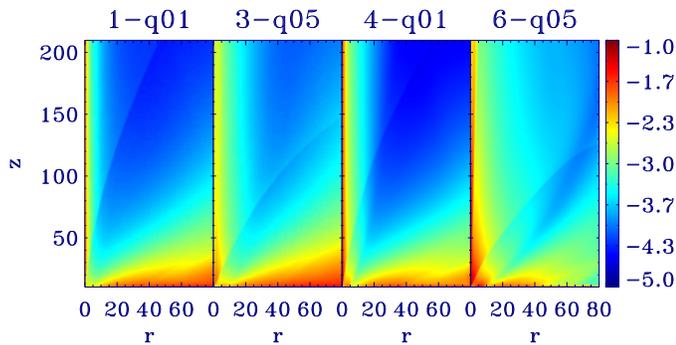}}
  \caption{
    Logarithm of the density for the final numerical solutions of models 1-q01,
    3-q05, 4-q01, 6-q05 (left to right).
    The two leftmost cases have a weaker stellar component compared to the two
    rightmost.
  }
  \label{fig:matching_surface}
\end{figure}

Fig.~\ref{fig:matching_surface} shows the logarithm of the final density of the
simulations carried out for models 1-q01, 3-q05, 4-q01, 6-q05 (left to right).
When the position of the matching surface is rooted closer to the disk rather
than the star, the shock seems to bend towards the midplane, confining the
unmodified ADO solution in a smaller domain.
This result indicates that as the spatial domination of the ASO solution becomes
larger, the ADO model controls a smaller portion of the box, thus forming the
shock closer to the disk.

\begin{figure}
  \resizebox{\hsize}{!}{\includegraphics{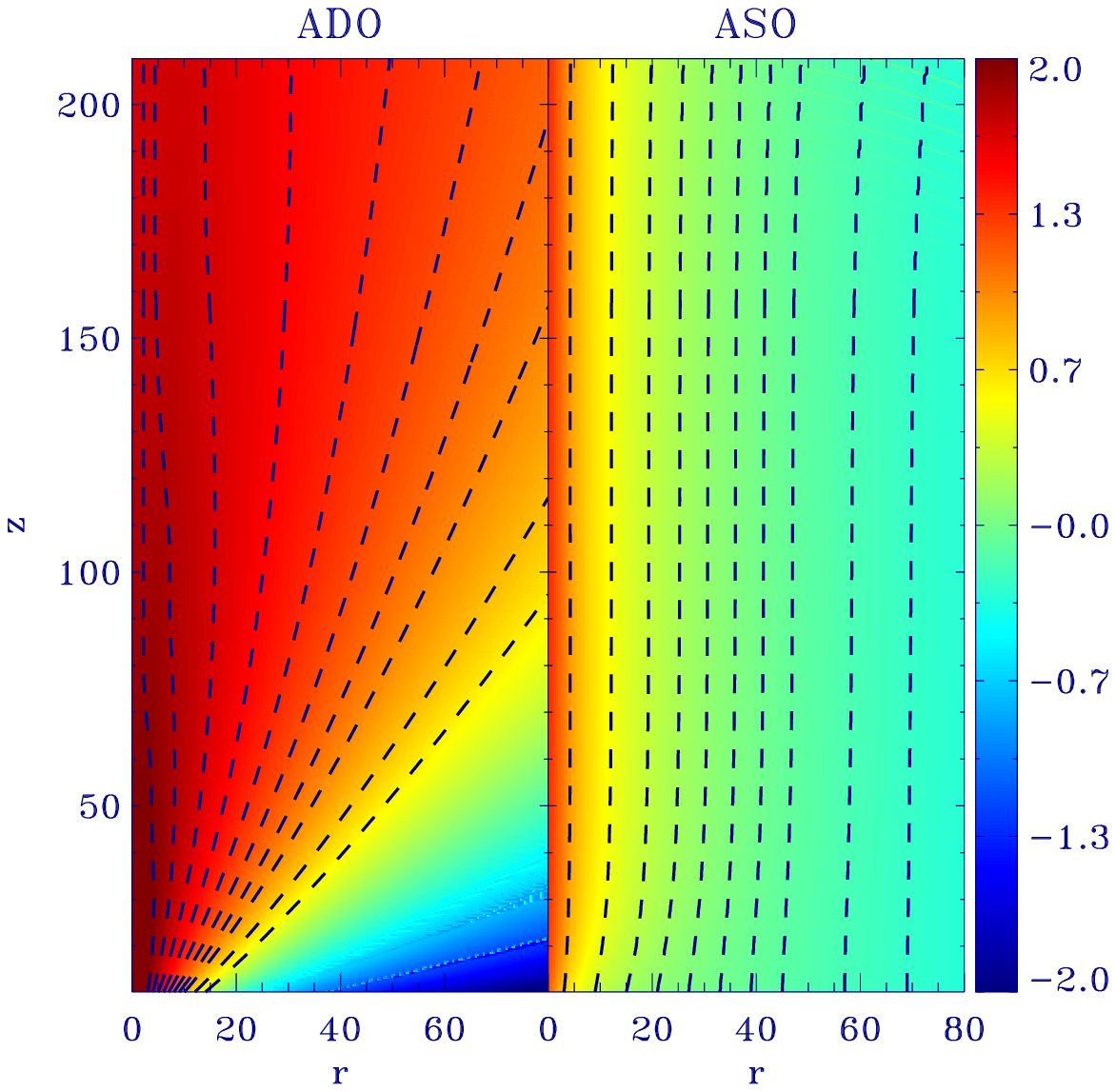}}
  \resizebox{\hsize}{!}{\includegraphics{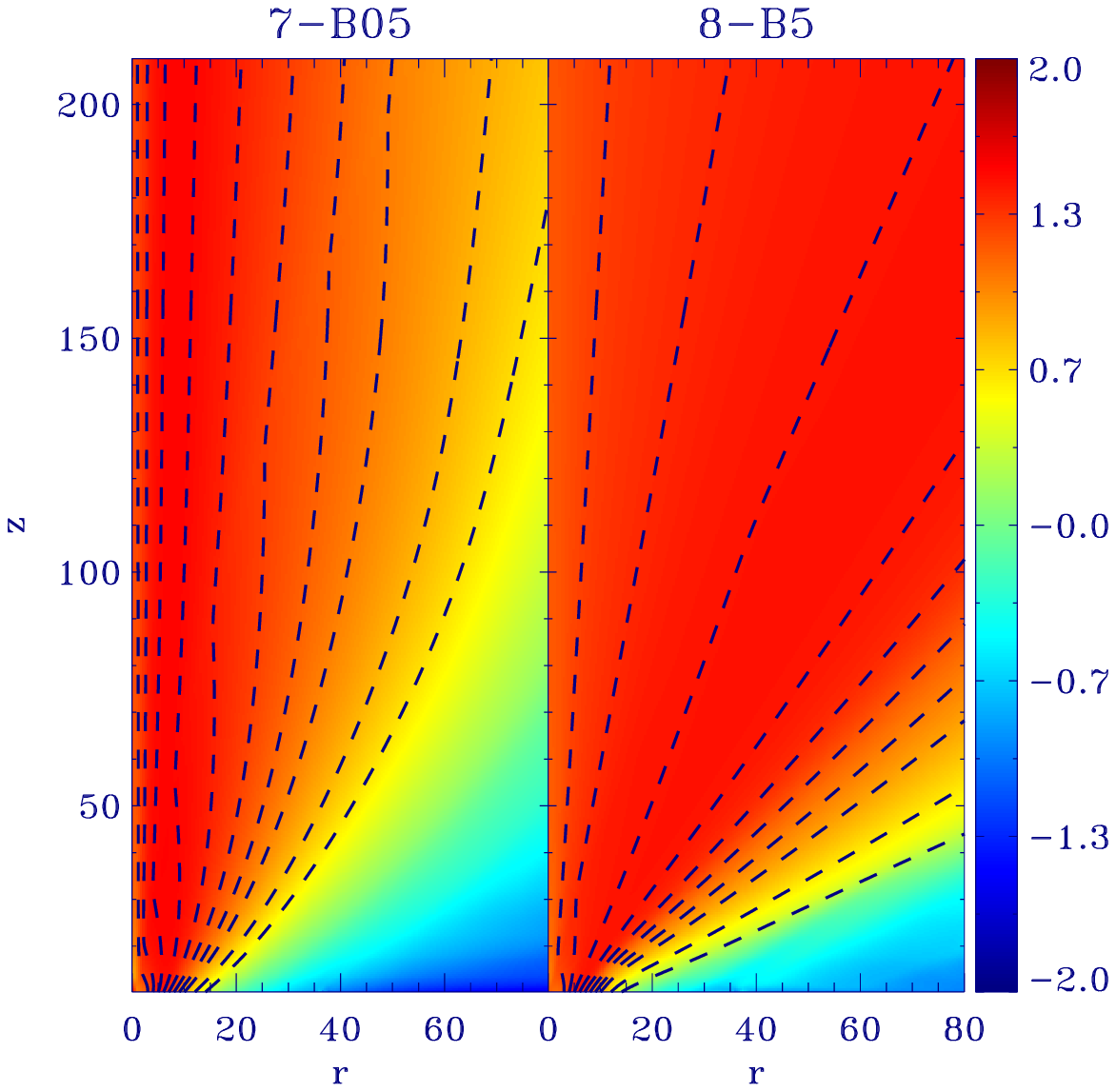}}
  \caption{
    Logarithmic poloidal velocity and streamlines (dashed lines) for the unmixed
    ADO (top left) and ASO (top right) models separately.
    In the lower panel, models 7-B05 ($\ell_B = 0.5$; left) and 8-B5
    ($\ell_B = 5$; right) are shown.
    The maximum values of the poloidal velocity of both of these two-component
    cases are $\sim 500\,$km$\,$s$^{-1}$, despite the misleading colorbar, which
    was accordingly chosen to match that of the top panel.
  }
  \label{fig:lambda_b}
\end{figure}

Recalling that models 1-q01 and 3-q05 have a weaker ASO contribution 
($\ell_B = 1$), compared to 4-q01 and 6-q05 ($\ell_B = 2$),
Fig.~\ref{fig:matching_surface} also suggests that the relative strength of the
ASO model does not seem to considerably affect the position of the shock,
although larger deviations are seen around the slow critical surface of the
initial ADO model.
The same result is derived from models 7-B05, 8-B5 (bottom of
Fig.~\ref{fig:lambda_b}) and 9-B10, where the location of the shock is found
farther from the axis, the larger the value of $\ell_B$.
Nevertheless, this might be related to the previous result since a large
$\ell_B$ also spatially reduces the contribution coming from the ADO model.

Furthermore, Fig.~\ref{fig:lambda_b} presents the logarithm of the poloidal
velocity and the streamlines (dashed lines) for the ADO and ASO solutions
separately, as well as for models 7-B05 ($\ell_B = 0.5$) and 8-B5
($\ell_B = 5$).
The left plot of the bottom panel suggests that for disk wind dominated jets,
the ADO solution is effectively collimating the central component.
However, we know that polytropically evolved ASO solutions become more
collimated and less dense than the non polytropic initial ASO models (Paper I).
So, it is rather difficult to disentangle the collimation due to the disk wind
and that due to the change in energetics.

Moreover, increasing by one order of magnitude the contribution of the ASO
model, the streamlines take an almost radial geometry (lower panel, right plot
of Fig.~\ref{fig:lambda_b}).
A similar result was obtained by Meliani et al. (\cite{Mel06}) when the mass
loss rate of the inner stellar wind becomes comparable to the disk mass loss
rate.
Although this might contradict the parallel flow structure seen in the right
plot of the top panel of Fig.~\ref{fig:lambda_b}, where the ASO solution is
plotted alone, we note that such a strong collimation comes from the linear
increase of $B_\phi$ (Fig.~\ref{fig:variables}, dashed line).
However, the two-component jet presents a more realistic distribution of
current, with a decreasing toroidal field at large distances
(Fig.~\ref{fig:variables}, crosses) and hence the hoop stress is not capable of
collimating the flow.

\begin{figure}
  \resizebox{\hsize}{!}{\includegraphics{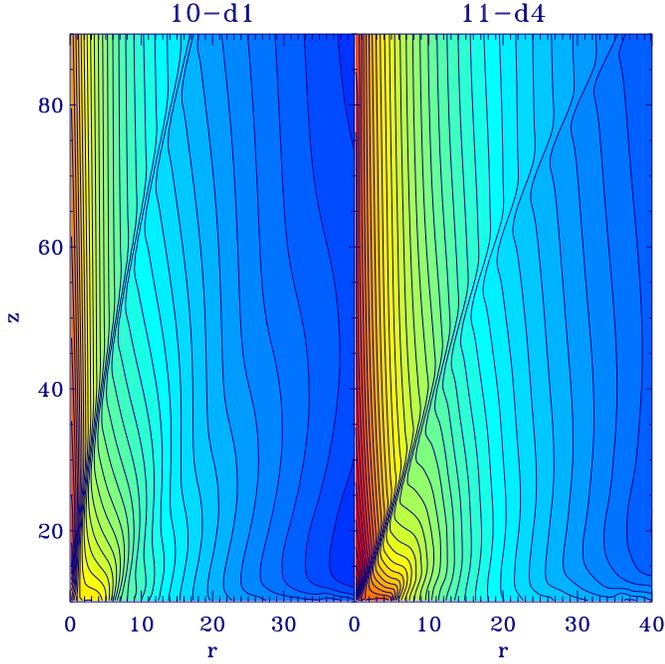}}
  \caption{
    Logarithmic pressure contours for model 10-d1 on the left and 11-d4 on the
    right.
  }
  \label{fig:steepness}
\end{figure}

Finally, we examine how the third free parameter, $d$, which defines the
steepness of the transition region, influences the final steady state reached by
the two-component jets.

The pressure contours of models 10-d1 and 11-d4 shown in
Fig.~\ref{fig:steepness} suggest that no matter how smoothly the variables
change from one solution to the other, the matching surface reaches the same
sort of structure at the end of the simulations.
On the other hand, the shock is affected more dramatically.
In the $d = 1$ case, it has a shape very similar to the one forming without the
presence of the ASO solution (see Paper I), with a polar angle of $\sim10^\circ$
as calculated close to the origin.
On the contrary, the shock intersects the axis with a wider polar angle
$\sim15^\circ$ in the $d = 4$ case.
Note that although the value of the parameter may change inside the
computational box, it is kept fixed at the lower boundary and hence influences
the evolution.

\subsection{Time variable stellar or X-type winds}

This last section is dedicated to the stability issues raised by a potential
time variability in the YSO's outflow.
We apply time dependency (Eq.~[\ref{eq:stellar_perturbation}] or
[\ref{eq:x_point_perturbation}]) either at the stellar wind's base or around the
X-point located at the interface between the stellar magnetosphere and the disk.
The two-component jet parameters adopted are identical to model 5-q02.

\begin{figure}
  \resizebox{\hsize}{!}{\includegraphics{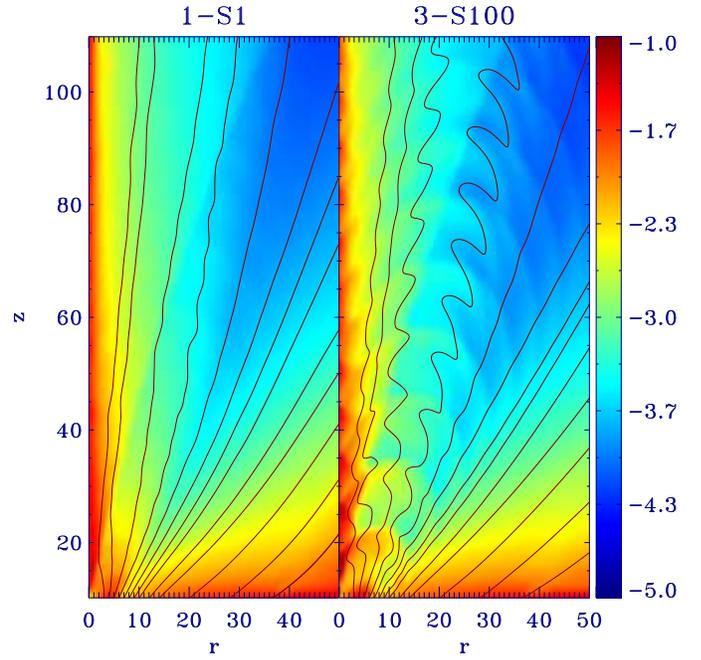}}
  \caption{
    Magnetic fieldlines (red) and logarithmic density contours at $t = 250$ for
    models 1-S1 (left) and 3-S100 (right).
  }
  \label{fig:aso_variability}
\end{figure}

High frequency velocity (or density) variations, associated with the Keplerian
rotation at roughly a stellar radius, seem to fade away on larger scales, as
shown on the left of Fig.~\ref{fig:aso_variability}.
The structure remains very close to the unperturbed model.
Two orders of magnitude lower frequency fluctuations, result in stronger
gradients along the flow, as seen in the right panel of
Fig.~\ref{fig:aso_variability}.
Considering that the velocity varies by $\pm50\%$ of its initial value, it is
surprising how well the two-component jet structure is retained.
Despite the ``wiggling'' of the magnetic field, the same flow features are found
as in the unperturbed cases.

\begin{figure}
  \resizebox{\hsize}{!}{\includegraphics{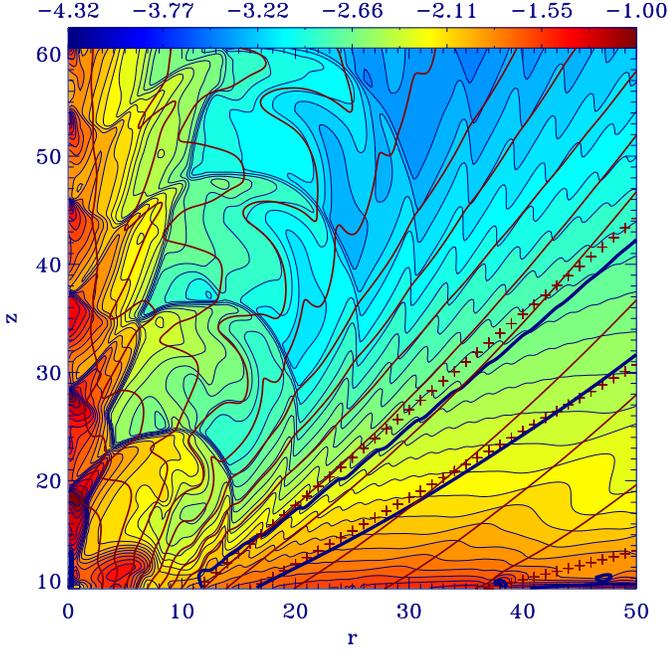}}
  \caption{
    Logarithmic density contours (thin blue lines) and the magnetic field
    (red lines) for model 3-S10$^2$ at $t = 50$.
    The red crosses and the thick blue lines denote the critical poloidal
    velocity surfaces of the ADO solution and those of the final equilibrium,
    respectively.
    This plot is equivalent to Fig.~\ref{fig:critical_surfaces}.
  }
  \label{fig:crit_surf_var}
\end{figure}

Fig.~\ref{fig:crit_surf_var} displays a plot equivalent to
Fig.~\ref{fig:critical_surfaces} in order to understand how the shock and
critical surfaces change in the time-variable stellar wind case.
The picture is very similar, apart from the perturbations seen in the density
throughout the computational box.
The poloidal critical surfaces show the same behavior as in the unperturbed
models and the weak steady shock is still present, being slightly curved locally
as the fluctuations propagate.

\begin{figure}
  \resizebox{\hsize}{!}{\includegraphics{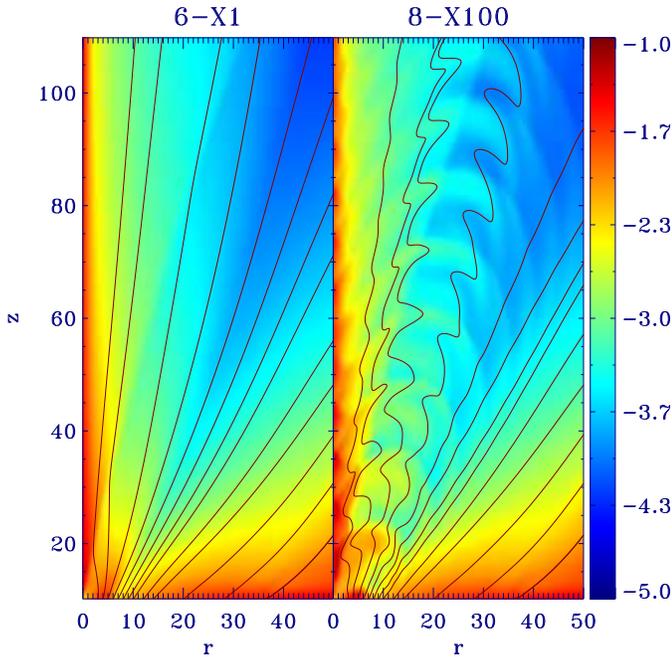}}
  \caption{
    Logarithm of the density along with magnetic fieldlines (red lines) at
    $t = 50$ for models 6-X1 and 8-X10$^2$.
  }
  \label{fig:x_variability}
\end{figure}

Analogous results are derived by the models where time variability is enforced
at the density and velocity of the outflow around the X-point.
The momentum changes periodically by an order of magnitude.
However, at the ASO and ADO dominated regions, the wind characteristics do not
seem to be affected, especially in the high frequency variability case (model
6-X1, plot of left panel of Fig.~\ref{fig:x_variability}).
On the other hand, more evident structures are produced in the 100 times lower
frequency fluctuations, still without destroying the basic pattern (model
8-X10$^2$, plot of right panel Fig.~\ref{fig:x_variability}).
This behaviour is similar to the stellar wind variability, but with a lesser
degree of collimation.

\begin{figure}
  \resizebox{\hsize}{!}{\includegraphics{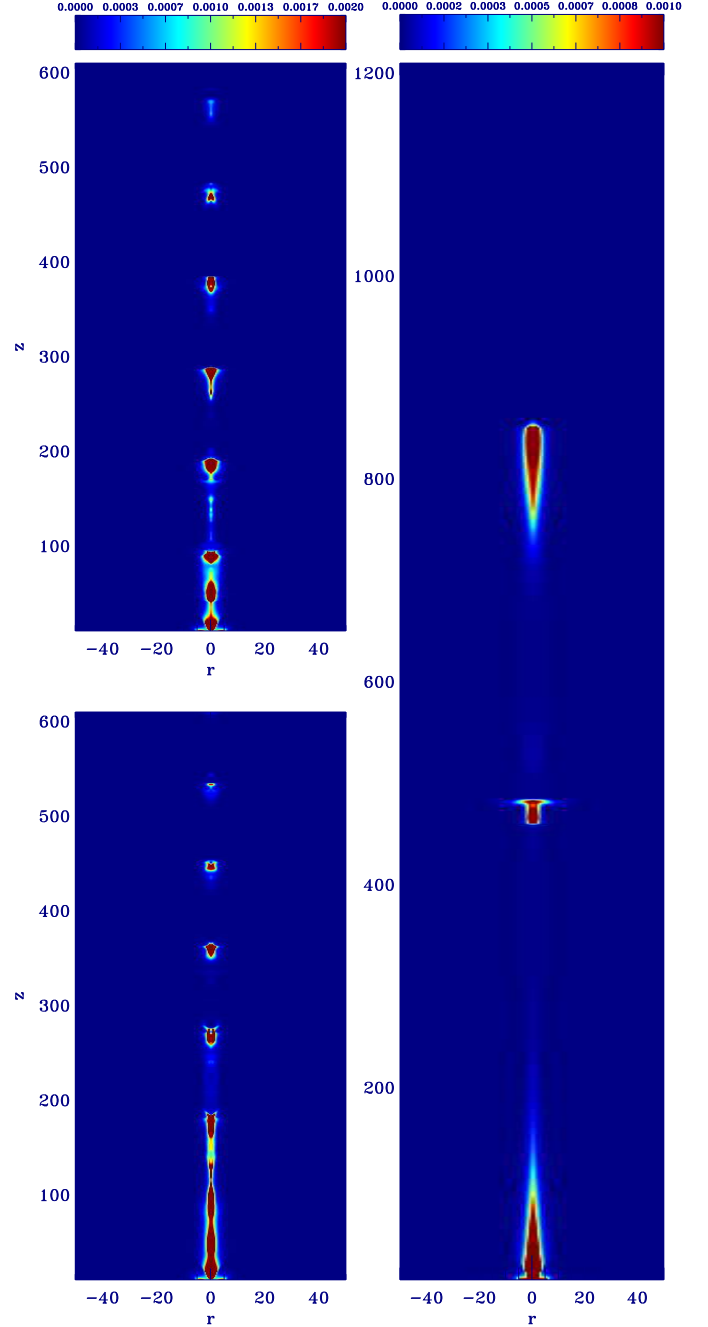}}
  \caption{
    The quantity $\rho^2\sqrt{T}$, which is roughly related with emissivity, is
    plotted for the low frequency variability models examined in large scales.
    In the left, models 4a-S10$^3$ (top) and 4b-S10$^3$ (bottom) are shown,
    whereas in the right model 5-S10$^4$ is displayed.
    Note that $\max(\rho^2\sqrt{T}) = 5.39\times10^{-2}$ for model 4a-S10$^3$,
    $\max(\rho^2\sqrt{T}) = 3.97\times10^{-2}$ for model 4b-S10$^3$ and
    $\max(\rho^2\sqrt{T}) = 7.49\times10^{-3}$ for model 5-S10$^4$.
    However, the colorbars use a lower maximum value in order to enhance the
    displayed features.
    The length code unit corresponds to $1\,$AU.
  }
  \label{fig:low_frequency}
\end{figure}

In order to see how the low frequency variability can affect the jet far away
from the launching region, we plot the quantity $\rho^2\sqrt{T}$ for models
4a-S10$^3$ (top left), 4b-S10$^3$ (bottom left) and 5-S10$^4$ (right) in
Fig.~\ref{fig:low_frequency}.
Close to the base, the numerical solutions remain close to the initial ADO and
ASO models.
However, at higher altitudes, the fluctuations create knot-like structures.
It is evident that such models can be associated with some jet variability.
We have checked that both stellar and X-wind type pulsations produce very
similar structures far away from the disk-star system.
This was expected, since kAU scales cannot distinguish the ejections coming from
within $1\,$AU.
The regular knot spacing observed in the jet of HH30 ($\sim100\,$AU, Bacciotti
et al. \cite{Bac99}) can be reasonably compared with our models 4a-S10$^3$ and
4b-S10$^3$, with a structure periodicity of $\sim1\,$yr.
Model 5-S10$^4$, with a periodicity of $10\,$yrs, could be associated with the
knots detected in the jet of HH34 where the condensation spacing is
$\sim1000\,$AU (Cohen \& Jones \cite{Coh87}).
Nevertheless, in this case there is a gap between the blobs and the star,
suggesting a contribution of other processes to the knot formation.
Note that the time scales of such fluctuations also correspond to typical
stellar variabilities (e.g. the $11\,$y period of the solar cycle).

\begin{figure}
  \resizebox{\hsize}{!}{\includegraphics{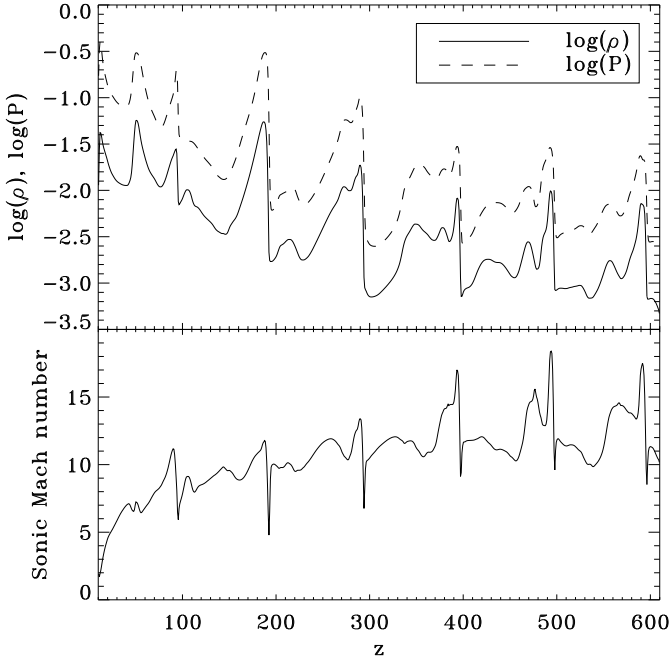}}
  \caption{
    Density, pressure and sonic Mach number plotted along the jet axis for model
    4a-S10$^3$.
  }
  \label{fig:knots}
\end{figure}

Finally, Fig.~\ref{fig:knots} provides the proof that these knot-like structures
are in fact shocks.
The top panel displays the periodic density and pressure jumps along the jet
axis, with the change being approximately an order of magnitude for both.
Note that close to the base the discontinuities are not yet well developed.
We also remark that these shocks are stronger than that associated with the FMSS
(see Fig.~\ref{fig:shock}).
The lower panel reports the sonic Mach number as a function of $z$.
Its mean value of the background flow is $\sim10$, in good accordance with YSO
jet observations.
The shocks propagate faster by $\sim50\%$, as expected in agreement with the
inflow time variability.
Although Fig.~\ref{fig:knots} suggests that the flow values converge to a
similar periodic structure, a larger computational box is needed to verify such
an argument.
In a future study, we plan to apply radiation cooling effects to these time
variable models, effectively producing realistic emission maps to be compared to
real data.

\section{Summary and conclusions}
  \label{sec:conclusions}

In this work, we have constructed two-component numerical jet models by properly
combining two well studied analytical solutions, each one describing separately
a disk wind (ADO) and a stellar outflow (ASO).
We have investigated the features of the time evolution and the characteristics
of the final outcome of the simulations as a function of the two-component jet
parameters and the enforced time variability.
Although the detailed launching mechanisms of each component are not taken into
account, the two-component jet models presented here seem able to capture the
dynamics and describe a variety of interesting scenarios.

The main conclusions of this work are the following:
\begin{itemize}
  \item
  The two-component jet models show remarkable stability and always reach a well
  defined steady state.
  This result is robust despite the fact that the two solutions have orthogonal
  symmetries, different geometry and different physics (i.e.
  launching mechanisms).
  In addition, the conclusion holds true independently of the choice of the
  parameters and even in the cases where time variability is enforced at the
  stellar wind's base or around the X-point.
  Therefore, the analytical solutions provide solid foundations for realistic
  two-component jet scenarios.
  \item
  The system remains close to the initial analytical solutions.
  In particular, the disk wind dominated regions are barely changed in the
  presence of the stellar outflow, with the exception of the slow magnetosonic
  regions.
  On the other hand, the central component is self consistently modified due to
  the assumption of a polytropic equation of state and because of the effective
  collimation caused by the surrounding disk wind.
  This implies that specific YSO systems can be addressed more accurately by
  constructing analytical outflow solutions with the desirable characteristics,
  before merging them into a two-component regime.
  \item
  A shock manifests during the time evolution, preventing any information from
  the downstream domain from reaching the base of the outflow.
  This separatrix causally disconnects the two-component jet from its launching
  regions, although initially there is no such ``horizon'' present in the
  computational box.
  The initial ASO solution does not exhibit any modified fast separatrix (Sauty
  et al. \cite{Sau02}), whereas despite the existence of the FMSS in the initial
  ADO model (at small polar angles), it is effectively replaced by the stellar
  wind in the initial setup.
  Nonetheless, the final equilibria reached by the numerical models show the
  formation of a weak shock corresponding to such a surface, causally
  disconnecting the acceleration regions from the jet propagation physics and
  subsequent interaction with the outer medium.
  \item
  We may address various two-component jet scenarios, by means of two parameters
  controlling the relative contribution of each component, $\ell_B$, and the
  time variability function, $f(t)$.
  With the former, we can smoothly switch the physics from a totally
  magneto-centrifugal wind to a pressure driven jet.
  With the latter, flow fluctuations are introduced, producing knot-like
  structures on large scales that are quantitatively similar to HH30 and HH34
  observations.
\end{itemize}

Thus, most of the technical part concerning two-component jets, e.g. 2.5D
stability, steady states, parameter study, time variability etc., is now
available, providing us with all the necessary ingredients to address YSO jets.
With a) the proper analytical solutions, i.e. desirable lever arm, mass loss
rate etc., b) the correct choice of the mixing parameters and c) an enforced
time variability that effectively produces knot structures, we are now ready to
qualitatively study different and realistic scenarios, address observed jet
properties and ultimately understand the various outflow phases of specific T
Tauri stars.
However, such applications and comparison with relevant observational data is
beyond the scope of this paper and will be presented in a future work.

\begin{acknowledgements}

We acknowledge V. Cayatte and the rest of the group in LUTh for fruitful
discussions, and an anonymous referee for helpful comments and suggestions that
resulted in a better presentation of this work.
We would also like to thank Capt. D. Kalogeras whose support during the revision
of this paper prevented a delay of several months.
The present work was supported in part by the European Community's Marie Curie
Actions - Human Resource and Mobility within the JETSET (Jet Simulations,
Experiments and Theory) network under contract MRTN-CT-2004 005592 and in part
by the HPC-EUROPA++ project (project number: 211437), with the support of the
European Community - Research Infrastructure Action of the FP7 ``Coordination
and support action'' Programme.

\end{acknowledgements}

\appendix

\section{The self-similar outflow formulation}
  \label{app:dependencies}

Axisymmetry, steady state and self-similarity assumptions simplify the ideal MHD
equations to a set of coupled ODEs in spherical coordinates.
These equations are solved numerically, providing the values of some key
functions for each model.

For the ADO solution (radially self-similar), the physical variables are
provided in terms of the tabulated key functions $G_D(\theta)$, $M_D(\theta)$
and $\psi_D(\theta)$:
\[
  \rho_D = \rho_{D*}\alpha_D^{x - 3/2}\frac{1}{M_D^2} \,,
  \quad
  P_D = P_{D*}\alpha_D^{x - 2}\frac{1}{M_D^{2\gamma}} \,,
\]\[
  \vec V_{Dp} = -V_{D*}\alpha_D^{-1/4}\frac{M_D^2}{G_D^2}
    \frac{\sin\theta}{\cos(\psi_D + \theta)}
    \left(\cos\psi_D \hat r + \sin\psi_D \hat z\right) \,,
\]\[
  V_{D\phi} = V_{D*}\lambda\alpha_D^{-1/4}
    \frac{G_D^2 - M_D^2}{G_D(1 - M_D^2)} \,,
\]\[
  \vec B_{Dp} = - B_{D*}\alpha_D^{x/2-1}\frac{1}{G_D^2}
    \frac{\sin\theta}{\cos(\psi_D + \theta)}
    \left(\cos\psi_D \hat r + \sin\psi_D \hat z\right) \,,
\]\[
  B_{D\phi} = -B_{D*}\lambda\alpha_D^{x/2-1}\frac{1 - G_D^2}{G_D(1 - M_D^2)} \,,
\]
where $p$ denotes the poloidal component.

The ASO solution (meridionally self-similar) is described with the help of the
key functions $G_S(R)$, $M_S(R)$, $F_S(R)$ and $\Pi_S(R)$:
\[
  \rho_S = \rho_{S*}\frac{1}{M_S^2}(1 + \delta\alpha_S) \,,
  \quad
  P_S = P_{S*}\Pi_S (1 + \kappa\alpha_S) \,,
\]\[
  V_{Sr} = V_{S*}\frac{M_S^2}{G_S^2} 
    \frac{\sin\theta \cos\theta}{\sqrt{1 + \delta\alpha_S}}
    \left(1 - \frac{F_S}{2}\right) \,,
\]\[
  V_{Sz} = V_{S*}\frac{M_S^2}{G_S^2}\frac{1}{\sqrt{1 + \delta\alpha_S}}
    \left(\cos^2\theta + \sin^2\theta\frac{F_S}{2}\right) \,,
\]\[
  V_{S\phi} = V_{S*}\lambda'\alpha_S^{1/2}\frac{G_S^2 - M_S^2}{G_S(1 - M_S^2)}
    \frac{1}{\sqrt{1 + \delta\alpha_S}} \,,
\]\[
  B_{Sr} = B_{S*}\frac{\sin\theta \cos\theta}{G_S^2}
  \left(1 - \frac{F_S}{2}\right) \,,
\]\[
  B_{Sz} = B_{S*}\frac{1}{G_S^2}
    \left(\cos^2\theta + \sin^2\theta\frac{F_S}{2}\right) \,,
\]\[
  B_{S\phi} = -B_{S*}\lambda'\alpha_S^{1/2}\frac{1 - G_S^2}{G_S(1 - M_S^2)} \,.
\]

\end{document}